\documentclass[CEJP,DVI]{cej} 
\usepackage{layout}
\usepackage{amsmath}
\usepackage{dsfont}
\usepackage{textcomp}
\usepackage{hyperref}

\newcommand\CC{{\mathds{C}}}
\newcommand\RR{{\mathds{R}}}
\newcommand\dd{{\mathrm{d}}}
\newcommand\ee{{\mathrm{e}}}
\newcommand\ii{{\mathrm{i}}}

\DeclareMathOperator{\Tr}{Tr}
\DeclareMathOperator{\2F1}{_2F_1}

\title{Long-time asymptotics of the long-range Emch-Radin model}

\articletype{Research Article}

\author{Michael Kastner\inst{1}$^,$\inst{2}\email{kastner@sun.ac.za}}

\institute{
     \inst{1} National Institute for Theoretical Physics (NITheP),\\
	Stellenbosch 7600, South Africa
     \inst{2} Institute of Theoretical Physics, University of Stellenbosch,\\
	 Stellenbosch 7600, South Africa
          }

\abstract{The long-time asymptotic behaviour is studied for a long-range variant of the Emch-Radin model of interacting spins. We derive upper and lower bounds on the expectation values of a class of observables. We prove analytically that the time scale at which the system relaxes to equilibrium diverges with the system size $N$, displaying quasistationary nonequilibrium behaviour. This finding implies that, for large enough $N$, equilibration will not be observed in an experiment of finite duration.}


\begin{document}
\maketitle


\section{Introduction}
Long-range interactions are commonly defined as pair interactions decaying like $r^{-\alpha}$, where $r$ is the distance between the interacting constituents, and the positive exponent $\alpha$ is smaller than or equal to the spatial dimension $d$ of the system. Examples of long-range interactions are common in nature, including gravitating masses, or Coulomb forces in the absence of screening. It has been known for several decades that long-range systems may show unconventional and sometimes surprising behaviour compared to the more familiar short-range interacting ones. For example, negative heat capacities were shown to exist in bounded self-gravitating gas spheres by Lynden-Bell and Wood \cite{LynWood68} in 1968, and Thirring explained this observation by relating it to the nonequivalence of microcanonical and canonical ensembles in the presence of long-range interactions \cite{Thirring70}. It was only recently in \cite{Kastner10,KastnerJSTAT10} that nonequivalence of ensemble has also been observed (theoretically) in long-range quantum spin systems like the one studied in the present article. Another particularly interesting phenomenon, going under the name of quasistationary states, has triggered intense activity, reviewed in \cite{CamDauxRuf09}, in the field of long-range interacting systems. The term {\em quasistationary}\/ is used to describe metastable states whose lifetime diverges with increasing system size $N$. The physical importance of quasistationary behaviour should be obvious: For a sufficiently large system, the transition from a quasistationary state to equilibrium takes place on a time scale that is larger than the experimentally accessible observation time. Hence, equilibrium properties will not be observed, and instead the statistical properties of the quasistationary regime are of interest.

Most of the studies of quasistationary states have focused on the Hamiltonian Mean-Field model \cite{AnRu95}, a toy model consisting of classical $XY$ spins (or plane rotators), each coupled to every other at equal strength. The typical scenario observed is that, for a suitable class of initial conditions, the total magnetization of this spin model rapidly relaxes to some quasistationary value different from its equilibrium value, remains there for a very long time, and finally decays to equilibrium (see Figure 33 in \cite{CamDauxRuf09}). More recently, quasistationary states have been observed in gravitational sheet models \cite{JoyceWorrakitpoonpon10} and their existence has been argued to be generic for a large class of classical long-range systems \cite{Gabrielli_etal10}. Virtually all finite-system results in the field have been obtained by numerical techniques, supplemented by analytical calculations in the $N\to\infty$ Vlasov continuum limit \cite{Barre_etal06,Antoniazzi_etal07,CamDauxRuf09,BouchetGuptaMukamel10,ChavanisBaldovinOrlandini11}.

In a recent Letter \cite{Kastner11}, the author has reported results on the dynamics of a long-range interacting version of the Emch-Radin model \cite{Emch66,Radin70}. This quantum mechanical spin model is particularly amenable to analytic calculations and is known to show relaxation to equilibrium in a non-Markovian way \cite{Emch66}. Unlike in the papers of Emch and Radin, long-range interaction are permitted, i.e.\ exponents $\alpha$ (determining the spatial decay of the interaction strength) which are smaller than the lattice dimension $d$. The results announced in \cite{Kastner11} include an analytic expression for the expectation value $\langle A\rangle(t)$ as a function of time $t$ for a certain class of observables $A$, as well as upper and lower bounds on $\langle A\rangle(t)$ in the thermodynamic limit of infinite system size. In particular, the lower bound establishes rigorously the occurrence of quasi-stationary behaviour in the long-range version of this model. The aim of the present article is twofold: (a) to provide a detailed derivation of the bounds on $\langle A\rangle(t)$, and (b) to numerically investigate how the equilibration time scales with the system size $N$. 

The article is structured as follows: In Section \ref{s:EmchRadin}, the Emch-Radin model is introduced, generalizing also to long-range spin-spin interactions. In Section \ref{s:finite}, an analytic expression for the time evolution of the expectation value $\langle A\rangle(t)$ is given for a class of observables $A$ and initial states $\rho(0)$. This is a rather straightforward extension of Emch's analysis to the long-range case. Section \ref{s:tdL} is devoted to the study of $\langle A\rangle(t)$ in the thermodynamic limit of infinite system size. In Section \ref{s:upperbound}, an upper bound on $\langle A\rangle(t)$ is constructed. For exponents $0\leq\alpha<1$, the lower bound derived in Section \ref{s:lowerbound} establishes that, in the thermodynamic limit, the system is trapped for all times at its initial state and relaxation to equilibrium is never observed. In Section \ref{s:rescaledtime} we perform a numerical analysis of the exact formula for $\langle A\rangle(t)$, studying in particular the scaling behaviour of the equilibration time scale $T_0$ with the system size. We find a power law $T_0\propto N^q$ with some exponent $q$, in agreement with what is predicted by the upper bound derived in Section \ref{s:upperbound}. The findings are summarized and discussed in Section \ref{s:discussion}.

\section{Emch-Radin model}
\label{s:EmchRadin}
Consider $N$ identical spin-$1/2$ particles, attached to the sites $\{1,\dotsc,N\}$ of a finite one-dimensional lattice with periodic boundary conditions.\footnote{Generalizations to higher-dimensional lattices are straightforward, but come at the expense of a more complicated notation. All results reported in this article generalize to dimensions $d\geq2$ with only insignificant modifications.} The quantum dynamics takes place on the Hilbert space
\begin{equation}
\mathcal{H}=\bigotimes_{i=1}^N \CC_i^2,
\end{equation}  
where the $\CC_i^2$ are identical replicas of the two-dimensional Hilbert space of a single spin-$1/2$ particle. The time evolution on $\mathcal{H}$ is generated by the Hamiltonian
\begin{equation}\label{e:Hamiltonian}
H_N=\mathcal{N}_\alpha\sum_{i=1}^N\sum_{j=1}^{N/2} \epsilon(j)\sigma_i^z \sigma_{i+j}^z - h\sum_{i=1}^N \sigma_i^z,
\end{equation}
with $\sigma^z$ denoting the $z$ component of the Pauli spin operator and $h\in\RR$ an external magnetic field in the $z$ direction. The index $i+j$ is to be considered modulo $N$ to account for periodic boundary conditions. The $\epsilon(j)$ are pair coupling constants depending on the distance $j$ of two spin operators on the lattice, and we assume
\begin{equation}\label{e:lime}
\lim_{j\to\infty}\epsilon(j)=0.
\end{equation}
A typical example we have in mind is the Dyson model \cite{Dyson69a} with algebraically decaying couplings, $\epsilon(j)=j^{-\alpha}$, but, in contrast to Dyson's work, exponents $\alpha$ smaller than 1 are also allowed. The explicit calculations reported in the present article are for such Dyson-type interactions, but generalizations are possible. For exponents $\alpha>1$, the interaction is absolutely summable,
\begin{equation}\label{e:l1}
\sum_{j=1}^\infty|\epsilon(j)|<\infty,
\end{equation}
and this case has already been treated in \cite{Emch66,Radin70,Wreszinski10}. Here we are interested in long-range interactions with $0\leq\alpha<1$ where the sum in \eqref{e:l1} diverges. In this case, the double sum in \eqref{e:Hamiltonian} gives rise to an infinite energy per spin $\langle H_N\rangle/N$ in the thermodynamic limit $N\to\infty$. To render this limit finite, the normalization
\begin{equation}\label{e:N}
\mathcal{N}_\alpha=\Biggl(2\sum_{j=1}^{N/2} \epsilon(j)\Biggr)^{-1}
\end{equation}
has been introduced in \eqref{e:Hamiltonian}. This normalization is a generalization of the so-called Kac prescription (introduced by Baker \cite{Baker61}) commonly used when studying long-range interacting systems. For large system sizes $N$, its asymptotic behaviour is of the form
\begin{equation}\label{e:Nasymptotic}
2\mathcal{N}_\alpha\sim
\begin{cases}
(1-\alpha)2^{1-\alpha}N^{\alpha-1} &\text{for $0\leq\alpha<1$},\\
1/\ln N &\text{for $\alpha=1$},\\
1/\zeta(\alpha) &\text{for $\alpha>1$},
\end{cases}
\end{equation}
where $\zeta$ denotes the Riemann zeta function. Equation \eqref{e:Nasymptotic} implies that $\mathcal{N}_\alpha$ vanishes in the thermodynamic limit for exponents $\alpha\leq1$.

We consider observables of the type
\begin{equation}\label{e:A}
A(a_1,\dotsc,a_N)=\sum_{i=1}^N a_i \sigma_i^x
\end{equation}
with real coefficients $a_i$, i.e.\ observables which are linear combinations of the $x$ components $\sigma_i^x$ of the spin operators. None of these observables commutes with the Hamiltonian \eqref{e:Hamiltonian}, and hence there is a chance of observing relaxation to equilibrium when studying the time evolution of expectation values $\langle A\rangle(t)$. For technical reasons, the analysis is restricted to initial states (i.e.\ initial density operators) $\rho(0)$ which are diagonal in the $\sigma_i^x$ tensor product eigenbasis of $\mathcal{H}$. This assumption simplifies the calculations but is not expected to be essential for the findings reported.

The Emch-Radin model, including the choices of observables and initial states, is inspired by induction decay experiments probing the pulsed magnetic resonance of nuclei in a CaF$_2$ crystal \cite{LoweNorberg57}. In these experiments, the decay of the $x$ component of the total magnetization $A(1,\dotsc,1)= \sum_{i=1}^N \sigma_i^x$ was measured, and it was found to be superimposed by oscillations (see Fig.\ 1 of \cite{LoweNorberg57}). Despite its simplifying assumptions, the model captures well both the decay and the beating observed experimentally. Moreover, the Emch-Radin model and its random generalizations \cite{Wreszinski10} have proved useful as paradigmatic models for which the approach to equilibrium in quantum systems can be studied analytically.

\section{Finite-system time evolution}
\label{s:finite}
\begin{figure}\center
\includegraphics[width=0.45\linewidth]{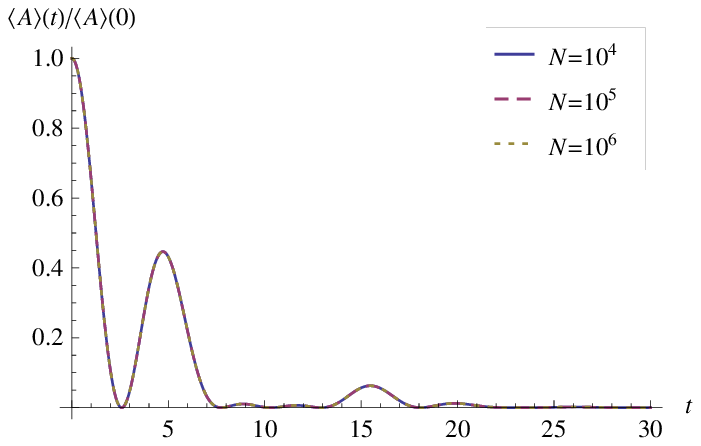}
\hspace{0.05\linewidth}
\includegraphics[width=0.45\linewidth]{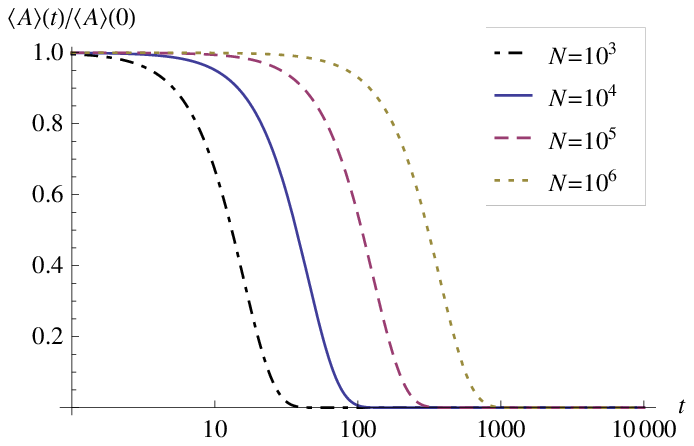}
\caption{\label{fig:decay}%
Time evolution of the expectation value \eqref{e:A_of_t} of an observable $A$ for magnetic field $h=0$ and various system sizes $N$. Left: For $\alpha=2$, an apparent decay is observed, superimposed by oscillations. The $N$ dependence of the time evolution is so weak that the curves for the various system sizes cannot be discerned in the plot. The behaviour for other $\alpha>1$ is qualitatively similar. Right: For $\alpha=1/2$, the expectation value again appears to be decaying, but on a time scale that depends strongly on the system size $N$ (note the logarithmic scale). Similar behaviour is observed for other values of $\alpha$ between zero and one.
}
\end{figure}

We study the time evolution, generated by the Hamiltonian $H_N$, of the expectation value of an observable $A$ of the form \eqref{e:A} with respect to the initial state $\rho(0)$,
\begin{equation}
\langle A\rangle(t)=\Tr\left[\ee^{-\ii H_N t} A \ee^{\ii H_N t}\rho(0)\right]
\end{equation}
(in units where $\hbar=1$). Performing the trace in the $\sigma_i^x$ eigenbasis of $\mathcal{H}$, the diagonal form of $\rho(0)$ implies that only the diagonal elements of the operator $\ee^{-\ii H_N t} A \ee^{\ii H_N t}$ are required, and it is this crucial ingredient which allows us to obtain, similar to the calculation in \cite{Emch66}, the exact result
\begin{equation}\label{e:A_of_t}
\langle A\rangle(t)=\langle A\rangle(0)\cos(2ht)\prod_{j=1}^{N/2} \cos^2[2\mathcal{N}_\alpha\epsilon(j)t].
\end{equation}
In comparison with the original Emch-Radin model, the important difference here is the explicit $N$ dependence of the argument of the cosine through the normalization $\mathcal{N}_\alpha$. Regarding the approach to equilibrium, the Larmor precession $\cos(2ht)$ is not relevant and we set $h=0$ in the following. The behaviour of \eqref{e:A_of_t} is plotted for exponents $\alpha=2$ and $\alpha=1/2$ and various system sizes $N$ in Fig.\ \ref{fig:decay}, and in all cases the expectation value of $A$ appears to be decaying in time to the microcanonical ensemble average $\langle A\rangle_\text{mic}=0$.%
However, this decay is only apparent, as we can read off from \eqref{e:A_of_t} that $\langle A\rangle(t)$ is an almost periodic function in time for all finite $N$, and recurrences (or Loschmidt echos) will therefore occur on much longer time scales than shown. To possibly observe true equilibration, we have to invoke, as often in statistical physics, the idealizing concept of the thermodynamic limit.

\section{Thermodynamic limit}
\label{s:tdL}
In the thermodynamic limit, recurrence times may diverge and an relaxation to equilibrium may take place. An analysis of the infinite system dynamics is most rigorously done in a $\ast$-algebraic language \cite{Radin70,Wreszinski10}, but it essentially boils down to discussing the large-$N$ limit of the product in \eqref{e:A_of_t}, which is the content of the present section. To keep the presentation simple, the following derivations are done for algebraically decaying interaction strengths, $\epsilon(j)=j^{-\alpha}$ with some positive exponent $\alpha$, but generalizations are possible.

\subsection{Upper bound on $\langle A\rangle(t)$}
\label{s:upperbound}
The aim of this section is to construct, in the thermodynamic limit $N\to\infty$, an upper bound on the product
\begin{equation}\label{e:P_N}
P_N=\prod_{j=1}^{N/2} \cos^2\left(\frac{2\mathcal{N}_\alpha t}{j^\alpha}\right)
\end{equation}
occurring in the expression for the expectation value $\langle A\rangle(t)$ in \eqref{e:A_of_t}. Since $|\cos(x)|\leq1$, we have
\begin{equation}\label{e:P2}
P_N\leq\prod_{j=1}^{N/2} \left|\cos\left(\frac{2\mathcal{N}_\alpha t}{j^\alpha}\right)\right|.
\end{equation}
Then, for given $t$ and $N$ there exists a finite integer
\begin{equation}
M_N(t)=\left\lceil \left|\frac{4\mathcal{N}_\alpha t}{\pi}\right|^{1/\alpha}\right\rceil,
\end{equation}
defined such that
\begin{equation}
\left|\frac{2\mathcal{N}_\alpha t}{j^\alpha}\right| < \frac{\pi}{2}\qquad\forall j\geq M_N(t).
\end{equation}
Hence we can split the product in \eqref{e:P2} in the following way,
\begin{equation}\label{e:P3}
P_N\leq \prod_{j=1}^{M_N(t)-1} \underbrace{\left|\cos\left(\frac{2\mathcal{N}_\alpha t}{j^\alpha}\right)\right|}_{\displaystyle\leq1}
\prod_{j=M_N(t)}^{N/2} \left|\cos\left(\frac{2\mathcal{N}_\alpha t}{j^\alpha}\right)\right|.
\end{equation}
For finite $N$ and some large $t$, it will happen that the second product in \eqref{e:P3} consists of no factors at all, resulting in a trivial (and fairly useless) upper bound $P_N\leq1$. In fact, this was to be expected, since the unitary, quasi-periodic dynamics of finite systems gives rise to recurrences. But we are interested in the large-$N$ limit where, for a given $t$ and large enough $N$, the second product in \eqref{e:P3} will consist of a large number of factors. By virtue of the definition of $M_N(t)$, the arguments of the cosines in these factors are all between $-\pi/2$ and $\pi/2$, and for this range of arguments the cosine can be bounded by
\begin{equation}
|\cos(x)|\leq1-\frac{4x^2}{\pi^2}.
\end{equation}
From elementary properties of the exponential function it then follows that
\begin{equation}\label{e:expsum}
\begin{split}
\lim_{N\to\infty}P_N&\leq \lim_{N\to\infty}\prod_{j=M_N(t)}^{N/2} \left[1-\left(\frac{4\mathcal{N}_\alpha t}{\pi j^\alpha}\right)^2\right]\\
&\leq \lim_{N\to\infty}\prod_{j=M_N(t)}^{N/2} \exp\left[-\left(\frac{4\mathcal{N}_\alpha t}{\pi j^\alpha}\right)^2\right]
= \lim_{N\to\infty}\exp\Biggl[-\left(\frac{4\mathcal{N}_\alpha t}{\pi}\right)^2\sum_{j=M_N(t)}^{N/2} \frac{1}{j^{2\alpha}}\Biggr].
\end{split}
\end{equation}
Asymptotically for large $N$, the sum in \eqref{e:expsum} is given by\footnote{In equations \eqref{e:sumapprox} and \eqref{e:upperbound} we exclude for simplicity the case $\alpha=1/2$ which, as in \eqref{e:Nasymptotic}, would lead to corrections logarithmic in $N$.}
\begin{equation}\label{e:sumapprox}
\sum_{j=M_N(t)}^{N/2} \frac{1}{j^{2\alpha}} \sim\sum_{j=1}^{N/2} \frac{1}{j^{2\alpha}} \sim 
\begin{cases}
\frac{N^{1-2\alpha}}{1-2\alpha} & \text{for $0\leq\alpha<1/2$},\\
\zeta(2\alpha) & \text{for $1/2<\alpha\leq1$},
\end{cases}
\end{equation}
where the asymptotic formula \eqref{e:Nasymptotic} has been used in the last step. Inserting this result into \eqref{e:expsum}, a Gaussian in $t$ bound on $P_N$ is obtained which translates, via \eqref{e:A_of_t}, into a Gaussian upper bound on the expectation values of the observables $A$,
\begin{equation}\label{e:upperbound}
\lim_{N\to\infty}|\langle A\rangle(t)| \leq |\langle A\rangle(0)|\lim_{N\to\infty}
\begin{cases}
\exp\left[-\left(\frac{4(1-\alpha)}{2^\alpha \pi}\right)^2 \frac{1}{1-2\alpha}N^{-1}t^2\right] & \text{for $0\leq\alpha<1/2$},\\
\exp\left[-\left(\frac{4(1-\alpha)}{2^\alpha \pi}\right)^2 \zeta(2\alpha)N^{2\alpha-2}t^2\right] & \text{for $1/2<\alpha\leq1$}.
\end{cases}
\end{equation}
In both cases, and therefore for all $0\leq\alpha\leq1$, the $N$-dependent exponent goes to zero in the large-$N$ limit. Hence, for long-range interactions, the upper bound is constant in time, $\lim_{N\to\infty}|\langle A\rangle(t)| \leq |\langle A\rangle(0)|$, and no sign of relaxation to equilibrium can be inferred from the bound. In a way, this is not too surprising, as this finding is in agreement with the large-$N$ trend of the finite-system results plotted in Figure \ref{fig:decay}. A more detailed discussion of the bound \eqref{e:upperbound} will follow in Section \ref{s:rescaledtime}.

\subsection{Lower bound on $\langle A\rangle(t)$}
\label{s:lowerbound}
The finite-system results in Figure \ref{fig:decay} indicate that, for long-range interactions with exponents $0\leq\alpha\leq1$ and in the thermodynamic limit $N\to\infty$, the Emch-Radin model apparently fails to approach equilibrium. To prove that this is indeed the case, and that the failure to construct a nontrivial upper bound in Section \ref{s:upperbound} was not due to the crudeness of the bound, we will derive a lower bound on the product $P_N$ for the case of long-range interactions.

Starting from definition \eqref{e:P_N}, we can write
\begin{equation}
P_N=\prod_{j=1}^{N/2} \cos^2\left(\frac{2\mathcal{N}_\alpha t}{j^\alpha}\right)
= \exp\Biggl\{\sum_{j=1}^{N/2} \ln\left[\cos^2\left(\frac{2\mathcal{N}_\alpha t}{j^\alpha}\right)\right]\Biggr\}.
\end{equation}
For any given $t$ there exists an $N_0(t)$ such that
\begin{equation}
\left|\frac{2\mathcal{N}_\alpha t}{j^\alpha}\right|<1\qquad\forall N\geq N_0(t).
\end{equation}
Then we can use the inequality $\cos^2(x)\geq 1-x^2$ to obtain the bound
\begin{equation}
P_N\geq \exp\Biggl\{\sum_{j=1}^{N/2} \ln\left[1-\left(\frac{2\mathcal{N}_\alpha t}{j^\alpha}\right)^2\right]\Biggr\},
\end{equation}
valid for $N\geq N_0(t)$. In the thermodynamic limit we can replace the sum by an integral,
\begin{equation}\label{e:hypergeo}
\begin{split}
\lim_{N\to\infty}P_N&\geq \lim_{N\to\infty}\exp\Biggl\{ N\int_{1/N}^{1/2} \dd x \ln\left[1-\left(\frac{2\mathcal{N}_\alpha t}{N^\alpha x^\alpha}\right)^2\right]\Biggr\}\\
&= \lim_{N\to\infty}\exp\Biggl\{ 2Nx\alpha\left[1-\2F1\biggl(-\frac{1}{2\alpha},1;1-\frac{1}{2\alpha};\left(\frac{2\mathcal{N}_\alpha t}{N^\alpha x^\alpha}\right)^2\biggr) + \frac{x}{2\alpha}\ln\biggl(1-\left(\frac{2\mathcal{N}_\alpha t}{N^\alpha x^\alpha}\right)^2\biggr)\right]_{x=1/N}^{x=1/2}\Biggr\},
\end{split}
\end{equation}
where $\2F1$ denotes the Gauss hypergeometric function. The primitive in the second line of \eqref{e:hypergeo} was obtained with {\sc Mathematica}. Using the large-$N$ asymptotic equality
\begin{equation}
\frac{2\mathcal{N}_\alpha t}{N^\alpha} \sim \frac{(1-\alpha)2^{1-\alpha}t}{N}
\end{equation}
as obtained from \eqref{e:Nasymptotic} for $0\leq\alpha\leq1$, the argument of the exponential function in \eqref{e:hypergeo} can be rewritten as
\begin{multline}\label{e:exponent}
2\alpha\2F1\biggl(1,-\frac{1}{2\alpha};1-\frac{1}{2\alpha};\frac{(1-\alpha)^2 t^2}{4^{\alpha-1}N^{2-2\alpha}}\biggr)
-N\alpha\2F1\biggl(1,-\frac{1}{2\alpha};1-\frac{1}{2\alpha};\frac{4(1-\alpha)^2 t^2}{N^2}\biggr)\\
-2\alpha+N\alpha
+\underbrace{\frac{N}{2}\ln\left(1-\frac{4(1-\alpha)^2 t^2}{N^2}\right)}_{\displaystyle =\mathcal{O}(1/N)}
-\underbrace{\ln\left(1-\frac{(1-\alpha)^2 t^2}{4^{\alpha-1}N^{2-2\alpha}}\right)}_{\displaystyle =\mathcal{O}\bigl(N^{2\alpha-2}\bigr)}.
\end{multline}
From the Gauss series expansion (15.2.1 in \cite{NIST})
we obtain the large-$N$ expansions
\begin{align}
2\alpha\2F1\biggl(1,-\frac{1}{2\alpha};1-\frac{1}{2\alpha};\frac{(1-\alpha)^2 t^2}{4^{\alpha-1}N^{2-2\alpha}}\biggr)&=2\alpha+\mathcal{O}\left(N^{2\alpha-2}\right),\label{e:GaussSeries1}\\
N\alpha\2F1\biggl(1,-\frac{1}{2\alpha};1-\frac{1}{2\alpha};\frac{4(1-\alpha)^2 t^2}{N^2}\biggr)&=N\alpha+\mathcal{O}\left(N^{-1}\right),\label{e:GaussSeries2}
\end{align}
finding that the leading order of the overall exponent \eqref{e:exponent} is $\mathcal{O}\left(N^{-1}\right)$ or $\mathcal{O}\left(N^{2\alpha-2}\right)$, depending on which one is larger. This results in a lower bound
\begin{equation}
\lim_{N\to\infty}P_N\geq1,
\end{equation}
valid for exponents $0\leq\alpha\leq1$. Together with the upper bound \eqref{e:upperbound}, this proves that
\begin{equation}
\lim_{N\to\infty}\langle A\rangle(t)=\langle A\rangle(0).
\end{equation}
So indeed, as suggested by extrapolating the finite-system plots in Figure \ref{fig:decay}, the infinite system is trapped for all times in its initial state, showing no sign of relaxation to equilibrium. Having in mind physical systems of finite size, it seems more useful to state the result in the following way.
\begin{proposition}\label{prop1}
Consider the Ising-type Hamiltonian $H_N$ defined in \eqref{e:Hamiltonian} with power law decaying interactions $\epsilon(j)=j^{-\alpha}$ with $0\leq\alpha<1$. Consider further an observable $A$ of the type \eqref{e:A} and an initial state $\rho(0)$ being diagonal in the $\sigma_i^x$ tensor product eigenbasis of the underlying Hilbert space $\mathcal{H}$. Then for the expectation value $\langle A\rangle(t)$ of $A$ with respect to $\rho(t)$, the following holds true: For any fixed time $\tau$ and some small $\delta>0$, there is a finite $n_0(\tau)$ such that
\begin{equation}
\left|\langle A\rangle(t)-\langle A\rangle(0)\right|<\delta\qquad\text{$\forall t<\tau$, $N>n_0(\tau)$}.
\end{equation}
\end{proposition}
Interpreting this result in terms of an experiment, we can think of an experimental resolution $\delta$ for the measurement of $A$, and some duration $\tau$ of the experiment. Then the above proposition states that, within the experimental resolution and for a large enough system, no deviation of $\langle A\rangle(t)$ from its initial value can be observed for times $t\leq\tau$.

\subsection{Relaxation to equilibrium in rescaled time}
\label{s:rescaledtime}
Although for the long-range Emch-Radin model in the thermodynamic limit relaxation to equilibrium does not take place for any finite time $t$, the plots in Figure \ref{fig:decay} suggest that equilibration can be observed when considering a suitably rescaled time variable. The case of Curie-Weiss-type interactions, i.e.\ an exponent $\alpha=0$, can serve as a simple, exactly solvable example: Defining the rescaled time variable $\tau_0=t/\sqrt{N}$, the product $P_N$ in \eqref{e:P_N} can be written as
\begin{equation}\label{e:CW}
P_N=\exp\left\{N\ln\left[\cos^2\left(\frac{2\tau_0}{\sqrt{N}}\right)\right]\right\} = \exp\left\{-4\tau_0^2+\mathcal{O}(1/N)\right\},
\end{equation}
yielding a Gaussian relaxation to equilibrium in the large-$N$ limit. For general exponents $0<\alpha\leq1$, such a simple calculation is not feasible. However, comparing the upper bound \eqref{e:upperbound} with the corrections to the lower bound in \eqref{e:GaussSeries1} and \eqref{e:GaussSeries2}, we can read off that an $\alpha$-dependent rescaled time
\begin{equation}\label{e:talpha}
\tau_\alpha = \begin{cases}
tN^{-1/2} & \text{for $0\leq\alpha<1/2$},\\
tN^{\alpha-1} & \text{for $1/2<\alpha\leq1$},
\end{cases}
\end{equation}
leads to $N$-independent Gaussian bounds. This implies that the expectation value $\langle A\rangle(\tau_\alpha)$ displays a Gaussian decay to equilibrium when plotted as a function of $\tau_\alpha$, and the microcanonical expectation value is correctly reproduced in the limit of $\tau_\alpha\to\infty$,
\begin{equation}
\lim_{\tau_\alpha \to\infty}\lim_{N\to\infty}\langle A\rangle(\tau_\alpha)=0.
\end{equation}

\begin{figure}\center
\includegraphics[width=0.31\linewidth]{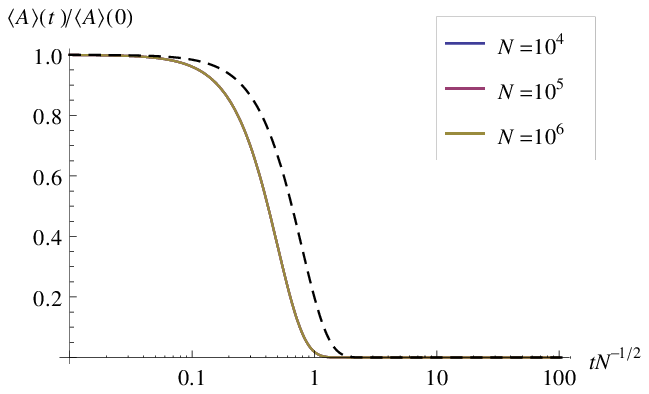}
\hspace{0.02\linewidth}
\includegraphics[width=0.31\linewidth]{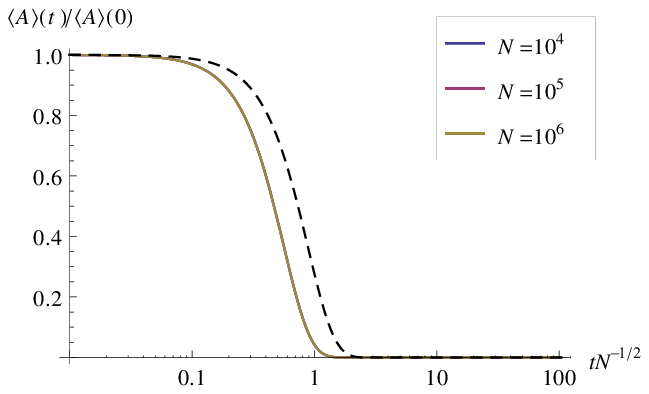}
\hspace{0.02\linewidth}
\includegraphics[width=0.31\linewidth]{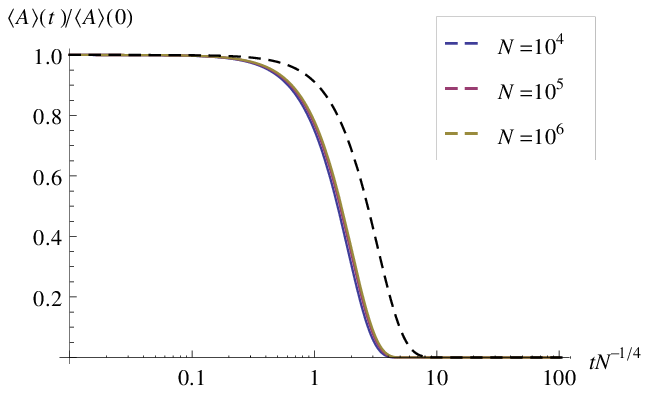}
\caption{\label{fig:collapse}%
Time evolution of the expectation value \eqref{e:A_of_t}, plotted as a function of (the logarithm of) $\tau_\alpha$ for various system sizes $N$. The graphs for different $N$ are hardly discernable in the plot, as they collapse onto each other. The three plots are for different values of the exponent $\alpha$ (from left to right: $\alpha=0$, $1/4$, $3/4$). The dashed lines show the upper bounds as given by Equation \eqref{e:upperbound}.
}
\end{figure}

We checked this prediction by numerically computing the product $P_N$ in \eqref{e:P_N} for various values of $N$, and plotting the result as a function of $\tau_\alpha$. As is shown in Figure \ref{fig:collapse}, the graphs of $P_N$ for different $N$ nicely collapse onto a single curve when plotted versus rescaled time $\tau_\alpha$. For comparison, the bounds \eqref{e:upperbound} are plotted as dashed lines in the same figure. This confirms that the time scale predicted by \eqref{e:talpha} is indeed the correct one. Rephrasing this result, we have found that the lifetime $T_0$ of quasi-stationary states in the long-range Emch-Radin model diverges with the system size like $T_0\sim N^q$ with an exponent $q=\min\{1/2,1-\alpha\}$.

\section{Discussion and outlook}
\label{s:discussion}
We have analytically studied the time evolution of long-range quantum spin models with Ising-type interactions \eqref{e:Hamiltonian} on one-dimensional lattices and for a certain class of initial states. Generalizations to higher-dimensional lattices are straightforward, but have not been discussed in this article. For finite system sizes $N$, almost periodic behaviour is observed with recurrence times that increase exponentially with $N$. In a rigorous sense, equilibration can therefore occur only in the thermodynamic limit of infinite lattice size. Whether this indeed happens depends on the asymptotic behaviour of the interaction strength $\epsilon(j)$ for large distances $j$ on the lattice. To analyze the long-time asymptotic behaviour of the expectation value $\langle A\rangle(t)$ in the thermodynamic limit, we constructed upper and lower bounds. For short-range interactions, i.e.\ interactions decaying faster than $Cj^{-\alpha}$ with some constant $C$ and exponent $\alpha>1$, it had already been proven in \cite{Radin70} that relaxation to equilibrium takes place as expected and the expectation value $\langle A\rangle(t)$ approaches the microcanonical ensemble average $\langle A\rangle_\text{mic}=0$ for large $t$.

For exponents $\alpha$ between zero and one, the interaction strength $\epsilon(j)$ is not absolutely summable and an $N$-dependent normalization $\mathcal{N}_\alpha$ is needed in \eqref{e:Hamiltonian} to render the energy per spin finite. In this case, the trivial (i.e.\ constant-in-time) upper bound $\langle A\rangle(t)\leq \langle A\rangle(0)$ is obtained, which does not permit any conclusions on whether or not $\langle A\rangle$ approaches its equilibrium value $\langle A\rangle_\text{mic}=0$. In order to get a more detailed picture of what happens in this case, we constructed a lower bound, finding that $\langle A\rangle(t)\geq \langle A\rangle(0)$ for exponents $0\leq\alpha<1$. Rephrasing this result, our Proposition \ref{prop1} asserts that, at a given instant of time and for large enough $N$, $\langle A\rangle(t)$ will be practically unchanged from its initial value $\langle A\rangle(0)$. This means that, for large enough systems, equilibration occurs on a time scale beyond the experimentally accessible one and will not be observed in practice. Despite the superficial similarities, such a behaviour is notably different from the failure of thermalization reported for near-integrable systems \cite{Kinoshita_etal06,Rigol09,Relano10}. Our results extend the concept of quasistationary states, previously observed and extensively studied in classical systems, into the realm of quantum mechanics. Moreover, we provide the first analytic proof of quasistationary behaviour from the microscopic time-evolution equations, as well as an expression for the $N$-dependence of the corresponding lifetime, $T_0\sim N^q$ with $q=\min\{1/2,1-\alpha\}$.

It is tempting to speculate that quasistationary behaviour, i.e.\ equilibration times that diverge for large system sizes $N$, might show up under much more general conditions, or even generically for long-range interacting quantum systems. An extension of our results to general anisotropic Heisenberg models might be a promising future project and, although an exact solution for the time evolution of $\langle A\rangle$ seems to be out of reach, the derivation of suitable bounds on $\langle A\rangle(t)$ might be feasible. Finally, the simultaneous presence of short- and long-range interactions might lead to equilibration taking place on two different time scales, possibly leading to more complex behaviour. The peculiarities of diverging relaxation times are expected to be reflected also in many applied aspects of quantum spin systems, including the study of quantum quenches \cite{Cramer_etal08,Polkovnikov_etal11}, decoherence, and many others.

As a concluding remark, a comment on the role of the normalization factor $\mathcal{N}_\alpha$ in the Hamiltonian \eqref{e:Hamiltonian} is in order. Since the Hamiltonian enters the time-evolution equation, it is not surprising to see that the system's dynamics inherits the $N$-dependence of $\mathcal{N}_\alpha$. But this is not the only source, as the term discussed in \eqref{e:sumapprox} introduces a further $N$-dependent contribution, and it is the interplay of the two terms which determines the nontrivial $N$-scaling of the relaxation time. So how should the prefactor $\mathcal{N}_\alpha$ in the Hamiltonian then be chosen? The answer to this question depends on the physical problem under investigation, and, depending on the situation, the following three choices may have their virtues:
\begin{enumerate}
\item The normalization factor $\mathcal{N}_\alpha$ as defined in \eqref{e:N} renders the energy an extensive quantity. As a result, energy and entropy have the same $N$-scaling in the large-$N$ limit. This is a necessary requirement for having competition between energetic and entropic effects, which in turn is crucial for the occurrence of thermodynamic phase transitions.
\item An $N$-dependent prefactor which leads to $N$-independent long-time asymptotic dynamics. This would eliminate quasi-stationary behaviour, but it is a very impractical choice, as it would require knowledge of the long-time asymptotic behaviour.
\item A prefactor independent of $N$, as this is what is realized in most physical systems. Such a choice is of course the most useful one when comparing experimental and theoretical results for systems of a given finite size, but it has the disadvantage of being unsuitable for the calculation of thermodynamic limits.
\end{enumerate}

\section*{Acknowledgments}
Financial support by the Incentive Funding for Rated Researchers programme of the National Research Foundation of South Africa is gratefully acknowledged.


\bibliography{Larnaca.bib}

\begin{thebibliography}{25}
\expandafter\ifx\csname natexlab\endcsname\relax\def\natexlab#1{#1}\fi
\expandafter\ifx\csname bibnamefont\endcsname\relax
  \def\bibnamefont#1{#1}\fi
\expandafter\ifx\csname bibfnamefont\endcsname\relax
  \def\bibfnamefont#1{#1}\fi
\expandafter\ifx\csname citenamefont\endcsname\relax
  \def\citenamefont#1{#1}\fi
\expandafter\ifx\csname url\endcsname\relax
  \def\url#1{\texttt{#1}}\fi
\expandafter\ifx\csname urlprefix\endcsname\relax\def\urlprefix{URL }\fi
\providecommand{\bibinfo}[2]{#2}
\providecommand{\eprint}[2][]{\url{#2}}

\bibitem[{\citenamefont{Lynden-Bell and Wood}(1968)}]{LynWood68}
\bibinfo{author}{\bibfnamefont{D.}~\bibnamefont{Lynden-Bell}} \bibnamefont{and}
  \bibinfo{author}{\bibfnamefont{R.}~\bibnamefont{Wood}},
  \bibinfo{journal}{Mon. Not. R. Astron. Soc.} \textbf{\bibinfo{volume}{138}},
  \bibinfo{pages}{495} (\bibinfo{year}{1968}).

\bibitem[{\citenamefont{Thirring}(1970)}]{Thirring70}
\bibinfo{author}{\bibfnamefont{W.}~\bibnamefont{Thirring}},
  \bibinfo{journal}{Z. Phys.} \textbf{\bibinfo{volume}{235}},
  \bibinfo{pages}{339} (\bibinfo{year}{1970}).

\bibitem[{\citenamefont{Kastner}(2010{\natexlab{a}})}]{Kastner10}
\bibinfo{author}{\bibfnamefont{M.}~\bibnamefont{Kastner}},
  \bibinfo{journal}{Phys. Rev. Lett.} \textbf{\bibinfo{volume}{104}},
  \bibinfo{pages}{240403} (\bibinfo{year}{2010}{\natexlab{a}}).

\bibitem[{\citenamefont{Kastner}(2010{\natexlab{b}})}]{KastnerJSTAT10}
\bibinfo{author}{\bibfnamefont{M.}~\bibnamefont{Kastner}}, \bibinfo{journal}{J.
  Stat. Mech.} \textbf{\bibinfo{volume}{2010}}, \bibinfo{pages}{P07006}
  (\bibinfo{year}{2010}{\natexlab{b}}).

\bibitem[{\citenamefont{Campa et~al.}(2009)\citenamefont{Campa, Dauxois, and
  Ruffo}}]{CamDauxRuf09}
\bibinfo{author}{\bibfnamefont{A.}~\bibnamefont{Campa}},
  \bibinfo{author}{\bibfnamefont{T.}~\bibnamefont{Dauxois}}, \bibnamefont{and}
  \bibinfo{author}{\bibfnamefont{S.}~\bibnamefont{Ruffo}},
  \bibinfo{journal}{Phys. Rep.} \textbf{\bibinfo{volume}{480}},
  \bibinfo{pages}{57} (\bibinfo{year}{2009}).

\bibitem[{\citenamefont{Antoni and Ruffo}(1995)}]{AnRu95}
\bibinfo{author}{\bibfnamefont{M.}~\bibnamefont{Antoni}} \bibnamefont{and}
  \bibinfo{author}{\bibfnamefont{S.}~\bibnamefont{Ruffo}},
  \bibinfo{journal}{Phys. Rev. E} \textbf{\bibinfo{volume}{52}},
  \bibinfo{pages}{2361} (\bibinfo{year}{1995}).

\bibitem[{\citenamefont{Joyce and
  Worrakitpoonpon}(2010)}]{JoyceWorrakitpoonpon10}
\bibinfo{author}{\bibfnamefont{M.}~\bibnamefont{Joyce}} \bibnamefont{and}
  \bibinfo{author}{\bibfnamefont{T.}~\bibnamefont{Worrakitpoonpon}},
  \bibinfo{journal}{J. Stat. Mech.} \textbf{\bibinfo{volume}{2010}},
  \bibinfo{pages}{P10012} (\bibinfo{year}{2010}).

\bibitem[{\citenamefont{Gabrielli et~al.}(2010)\citenamefont{Gabrielli, Joyce,
  and Marcos}}]{Gabrielli_etal10}
\bibinfo{author}{\bibfnamefont{A.}~\bibnamefont{Gabrielli}},
  \bibinfo{author}{\bibfnamefont{M.}~\bibnamefont{Joyce}}, \bibnamefont{and}
  \bibinfo{author}{\bibfnamefont{B.}~\bibnamefont{Marcos}},
  \bibinfo{journal}{Phys. Rev. Lett.} \textbf{\bibinfo{volume}{105}},
  \bibinfo{pages}{210602} (\bibinfo{year}{2010}).

\bibitem[{\citenamefont{Barr\'e et~al.}(2006)\citenamefont{Barr\'e, Bouchet,
  Dauxois, Ruffo, and Yamaguchi}}]{Barre_etal06}
\bibinfo{author}{\bibfnamefont{J.}~\bibnamefont{Barr\'e}},
  \bibinfo{author}{\bibfnamefont{F.}~\bibnamefont{Bouchet}},
  \bibinfo{author}{\bibfnamefont{T.}~\bibnamefont{Dauxois}},
  \bibinfo{author}{\bibfnamefont{S.}~\bibnamefont{Ruffo}}, \bibnamefont{and}
  \bibinfo{author}{\bibfnamefont{Y.~Y.} \bibnamefont{Yamaguchi}},
  \bibinfo{journal}{Physica A} \textbf{\bibinfo{volume}{365}},
  \bibinfo{pages}{177} (\bibinfo{year}{2006}).

\bibitem[{\citenamefont{Antoniazzi et~al.}(2007)\citenamefont{Antoniazzi,
  Fanelli, Barr\'e, Chavanis, Dauxois, and Ruffo}}]{Antoniazzi_etal07}
\bibinfo{author}{\bibfnamefont{A.}~\bibnamefont{Antoniazzi}},
  \bibinfo{author}{\bibfnamefont{D.}~\bibnamefont{Fanelli}},
  \bibinfo{author}{\bibfnamefont{J.}~\bibnamefont{Barr\'e}},
  \bibinfo{author}{\bibfnamefont{P.-H.} \bibnamefont{Chavanis}},
  \bibinfo{author}{\bibfnamefont{T.}~\bibnamefont{Dauxois}}, \bibnamefont{and}
  \bibinfo{author}{\bibfnamefont{S.}~\bibnamefont{Ruffo}},
  \bibinfo{journal}{Phys. Rev. E} \textbf{\bibinfo{volume}{75}},
  \bibinfo{pages}{011112} (\bibinfo{year}{2007}).

\bibitem[{\citenamefont{Bouchet et~al.}(2010)\citenamefont{Bouchet, Gupta, and
  Mukamel}}]{BouchetGuptaMukamel10}
\bibinfo{author}{\bibfnamefont{F.}~\bibnamefont{Bouchet}},
  \bibinfo{author}{\bibfnamefont{S.}~\bibnamefont{Gupta}}, \bibnamefont{and}
  \bibinfo{author}{\bibfnamefont{D.}~\bibnamefont{Mukamel}},
  \bibinfo{journal}{Physica A} \textbf{\bibinfo{volume}{389}},
  \bibinfo{pages}{4389} (\bibinfo{year}{2010}).

\bibitem[{\citenamefont{Chavanis et~al.}(2011)\citenamefont{Chavanis, Baldovin,
  and Orlandini}}]{ChavanisBaldovinOrlandini11}
\bibinfo{author}{\bibfnamefont{P.-H.} \bibnamefont{Chavanis}},
  \bibinfo{author}{\bibfnamefont{F.}~\bibnamefont{Baldovin}}, \bibnamefont{and}
  \bibinfo{author}{\bibfnamefont{E.}~\bibnamefont{Orlandini}},
  \bibinfo{journal}{Phys. Rev. E} \textbf{\bibinfo{volume}{83}},
  \bibinfo{pages}{040101} (\bibinfo{year}{2011}).

\bibitem[{\citenamefont{Kastner}(2011)}]{Kastner11}
\bibinfo{author}{\bibfnamefont{M.}~\bibnamefont{Kastner}},
  \bibinfo{journal}{Phys. Rev. Lett} \textbf{\bibinfo{volume}{106}},
  \bibinfo{pages}{130601} (\bibinfo{year}{2011}).

\bibitem[{\citenamefont{Emch}(1966)}]{Emch66}
\bibinfo{author}{\bibfnamefont{G.~G.} \bibnamefont{Emch}}, \bibinfo{journal}{J.
  Math. Phys. (N.Y.)} \textbf{\bibinfo{volume}{7}}, \bibinfo{pages}{1198}
  (\bibinfo{year}{1966}).

\bibitem[{\citenamefont{Radin}(1970)}]{Radin70}
\bibinfo{author}{\bibfnamefont{C.}~\bibnamefont{Radin}}, \bibinfo{journal}{J.
  Math. Phys. (N.Y.)} \textbf{\bibinfo{volume}{11}}, \bibinfo{pages}{2945}
  (\bibinfo{year}{1970}).

\bibitem[{\citenamefont{Dyson}(1969)}]{Dyson69a}
\bibinfo{author}{\bibfnamefont{F.~J.} \bibnamefont{Dyson}},
  \bibinfo{journal}{Commun. Math. Phys.} \textbf{\bibinfo{volume}{12}},
  \bibinfo{pages}{91} (\bibinfo{year}{1969}).

\bibitem[{\citenamefont{Wreszinski}(2010)}]{Wreszinski10}
\bibinfo{author}{\bibfnamefont{W.~F.} \bibnamefont{Wreszinski}},
  \bibinfo{journal}{J. Stat. Phys.} \textbf{\bibinfo{volume}{138}},
  \bibinfo{pages}{567} (\bibinfo{year}{2010}).

\bibitem[{\citenamefont{Baker}(1961)}]{Baker61}
\bibinfo{author}{\bibfnamefont{G.~A.} \bibnamefont{Baker}},
  \bibinfo{journal}{Phys. Rev.} \textbf{\bibinfo{volume}{122}},
  \bibinfo{pages}{1477} (\bibinfo{year}{1961}).

\bibitem[{\citenamefont{Lowe and Norberg}(1957)}]{LoweNorberg57}
\bibinfo{author}{\bibfnamefont{I.~J.} \bibnamefont{Lowe}} \bibnamefont{and}
  \bibinfo{author}{\bibfnamefont{R.~E.} \bibnamefont{Norberg}},
  \bibinfo{journal}{Phys. Rev.} \textbf{\bibinfo{volume}{107}},
  \bibinfo{pages}{46} (\bibinfo{year}{1957}).

\bibitem[{\citenamefont{Olver et~al.}(2010)\citenamefont{Olver, Lozier,
  Boisvert, and Clark}}]{NIST}
\bibinfo{editor}{\bibfnamefont{F.~W.~J.} \bibnamefont{Olver}},
  \bibinfo{editor}{\bibfnamefont{D.~W.} \bibnamefont{Lozier}},
  \bibinfo{editor}{\bibfnamefont{R.~F.} \bibnamefont{Boisvert}},
  \bibnamefont{and} \bibinfo{editor}{\bibfnamefont{C.~W.} \bibnamefont{Clark}},
  eds., \emph{\bibinfo{title}{{NIST} Handbook of Mathematical Functions}}
  (\bibinfo{publisher}{Cambridge University Press, Cambridge},
  \bibinfo{year}{2010}).

\bibitem[{\citenamefont{Kinoshita et~al.}(2006)\citenamefont{Kinoshita, Wenger,
  and Weiss}}]{Kinoshita_etal06}
\bibinfo{author}{\bibfnamefont{T.}~\bibnamefont{Kinoshita}},
  \bibinfo{author}{\bibfnamefont{T.}~\bibnamefont{Wenger}}, \bibnamefont{and}
  \bibinfo{author}{\bibfnamefont{D.~S.} \bibnamefont{Weiss}},
  \bibinfo{journal}{Nature (London)} \textbf{\bibinfo{volume}{440}},
  \bibinfo{pages}{900} (\bibinfo{year}{2006}).

\bibitem[{\citenamefont{Rigol}(2009)}]{Rigol09}
\bibinfo{author}{\bibfnamefont{M.}~\bibnamefont{Rigol}},
  \bibinfo{journal}{Phys. Rev. Lett.} \textbf{\bibinfo{volume}{103}},
  \bibinfo{pages}{100403} (\bibinfo{year}{2009}).

\bibitem[{\citenamefont{Rela{\~{n}}o}(2010)}]{Relano10}
\bibinfo{author}{\bibfnamefont{A.}~\bibnamefont{Rela{\~{n}}o}},
  \bibinfo{journal}{J. Stat. Mech.} \textbf{\bibinfo{volume}{2010}},
  \bibinfo{pages}{P07016} (\bibinfo{year}{2010}).

\bibitem[{\citenamefont{Cramer et~al.}(2008)\citenamefont{Cramer, Dawson,
  Eisert, and Osborne}}]{Cramer_etal08}
\bibinfo{author}{\bibfnamefont{M.}~\bibnamefont{Cramer}},
  \bibinfo{author}{\bibfnamefont{C.~M.} \bibnamefont{Dawson}},
  \bibinfo{author}{\bibfnamefont{J.}~\bibnamefont{Eisert}}, \bibnamefont{and}
  \bibinfo{author}{\bibfnamefont{T.~J.} \bibnamefont{Osborne}},
  \bibinfo{journal}{Phys. Rev. Lett.} \textbf{\bibinfo{volume}{100}},
  \bibinfo{pages}{030602} (\bibinfo{year}{2008}).

\bibitem[{\citenamefont{Polkovnikov et~al.}(2011)\citenamefont{Polkovnikov,
  Sengupta, Silva, and Vengalattore}}]{Polkovnikov_etal11}
\bibinfo{author}{\bibfnamefont{A.}~\bibnamefont{Polkovnikov}},
  \bibinfo{author}{\bibfnamefont{K.}~\bibnamefont{Sengupta}},
  \bibinfo{author}{\bibfnamefont{A.}~\bibnamefont{Silva}}, \bibnamefont{and}
  \bibinfo{author}{\bibfnamefont{M.}~\bibnamefont{Vengalattore}},
  \bibinfo{journal}{Rev. Mod. Phys.} \textbf{\bibinfo{volume}{83}},
  \bibinfo{pages}{863} (\bibinfo{year}{2011}).

\end{thebibliography}

\end{document}